\documentclass[12pt,a4paper]{article}
\usepackage{graphicx}
\usepackage{times}
\usepackage{natbib}
\usepackage{caption}
\usepackage{wrapfig,lipsum,booktabs}
\newcommand{\teff}{$T_{\scriptsize{\textrm{eff}}}$}
\newcommand{\logg}{$\log g$}
\newcommand{\feh}{[Fe/H]}

\title{\bf Low resolution spectroscopic investigation of Am stars using Automated method}
\author{Kaushal Sharma$^{1,3}$\thanks{kaushals@iucaa.in} , Santosh Joshi$^2$ and H. P. Singh$^3$\\
\vspace{1cm}\\
\normalsize $^1$ Inter-University Centre for Astronomy and Astrophysics, IUCAA, Pune - 411007, India\\
\normalsize $^2$ Aryabhatta Research Institute of Observational Sciences, Nainital- 263002, India\\
\normalsize $^3$ Department of Physics and Astrophysics, University of Delhi, Delhi - 110007, India}
\date{\mbox{}}
\begin{document}
\maketitle
\pagestyle{empty}
%
%
\def\bull{\vrule height .9ex width .8ex depth -.1ex}
\makeatletter
\def\ps@plain{\let\@mkboth\gobbletwo
\def\@oddhead{}\def\@oddfoot{\hfil\scriptsize\bull\quad
``First Belgo-Indian Network for Astronomy \& astrophysics (BINA) workshop'', held in Nainital (India), 15-18 November 2016 \quad\bull}%
\def\@evenhead{}\let\@evenfoot\@oddfoot}
\makeatother
%
%
\def\beginrefer{\section*{References}%
\begin{quotation}\mbox{}\par}
\def\refer#1\par{{\setlength{\parindent}{-\leftmargin}\indent#1\par}}
\def\endrefer{\end{quotation}}
%
%
{\noindent\small{\bf Abstract:}
Automated method of full spectrum fitting gives reliable estimates of stellar atmospheric parameters (\teff, \logg{} and \feh) for late A, F, G and early K type stars. Recently, the technique was further improved in the cooler regime and the validity range was extended up to M6\,-\,M7 spectral type( \teff{}\,$\sim\,2900$\,K). The present study aims to explore the application of this method on the low-resolution spectra of Am stars, a class of chemically peculiar (CP) stars, to examine its robustness for these objects. We use ULySS with MILES (Medium-resolution INT Library of Empirical Spectra) V2 spectral interpolator for parameter determination. Determined \teff{} and \logg{} are found to be in good agreement with those obtained from high-resolution spectroscopy.

}
%
%
\section{Introduction}

An avalanche of astronomical data is being generated from large ground and space based surveys, e.g. NASA's Hubble Space Telescope (HST), Sloan Digital Sky Survey (SDSS; York et al. (2000)) and European Space Agency's Gaia mission (Perryman et al. 2001). For processing and analysing these huge datasets, we need to develop automatic algorithms, which could be easily implemented in a parameter reduction pipeline at these survey programs for quick and reliable preliminary results.

Full spectrum fitting is one such method, which has been used for determining the reliable atmospheric parameters of F, G, K type stars (Wu et al. 2011, Prugniel et al. 2011) and recently for M type stars as well (Sharma et al. 2016). For investigating the potential of this technique for other stellar species, we test it on the low-resolution spectra of chemically peculiar stars.

Chemical peculiar (CP) stars are characterized by the abnormal abundances of chemical elements in their photosphere as compared to the normal stars of same spectral type. Their atmospheric parameters are crucial for understanding the physics governing the peculiarity, detailed abundance analysis and finding out their evolutionary status. However, there are some challenges in retrieving their parameters. Direct/semi-direct methods of temperature determination (such as IRFM) are difficult to apply as these require few physical properties (e.g. radius, angular diameter) to be accurately known in prior. Photometric calibrations, applicable for normal stars, can not be directly applied to these objects as they give systematically higher effective temperatures (Lipski \& St{\c e}pie{\'n}, 2008). Many high-resolution spectroscopic studies have also been performed to determine their parameters as well as detailed abundances (Joshi et al., 2012; Catanzaro et al., 2014), but this mode requires sophisticated instrumentation and involves more complex data analysis as compared to low-resolution spectroscopy. Low-resolution spectroscopy could be particularly useful in such cases where detailed abundances are not desired and only the atmospheric parameters are essential for the analysis.


In this study, we use the low resolution spectra of six CP1 stars, a subgroup of CP stars classically known as metallic line Am stars. 
To derive their atmospheric parameters, we use full spectrum fitting technique, employed in the ULySS package (Koleva et al., 2009). In Sec.~\ref{sec:observation}, we present the observing information and steps involved in the data pre-processing. Sec.~\ref{sec:parameters} describes the technique used to derive the atmospheric parameters of sample stars. Finally in Sec.~\ref{sec:discussion}, we briefly discuss the results obtained and the future scope of the current study.

\section{Observations and Data Reduction}
\label{sec:observation}

Sample stars are selected from the catalogue of the Nainital-Cape survey (Martinez et al., 2001; Joshi et al., 2003) and their spectral types, compiled from the literature, are presented in Table ~\ref{table:samples}. 

\begin{wraptable}{r}{6.0cm}
\caption{List of the sample stars. $^a$ Spectral types are taken from the Renson catalogue (Renson \& Manfroid, 2009). `dD' in the spectral class suggests that the star is a $\delta$ Del-type star. Spectral types are followed by elements with abnormal abundances.}\label{table:samples}
\begin{tabular}{lll}
\hline
S. No. & Star & Sp. Type$^a$ \\
\hline
1. & HD\,13038     & A4    \\
2. & HD\,13079     & F0    \\
3. & HD\,25515     & F0-dD \\
4. & HD\,98851     & F1 Sr \\
5. & HD\,102480    & F2 Sr \\
6. & HD\,113878    & A9 Sr \\
\hline
\end{tabular}
\end{wraptable}

\begin{figure}
 \begin{center}
\includegraphics{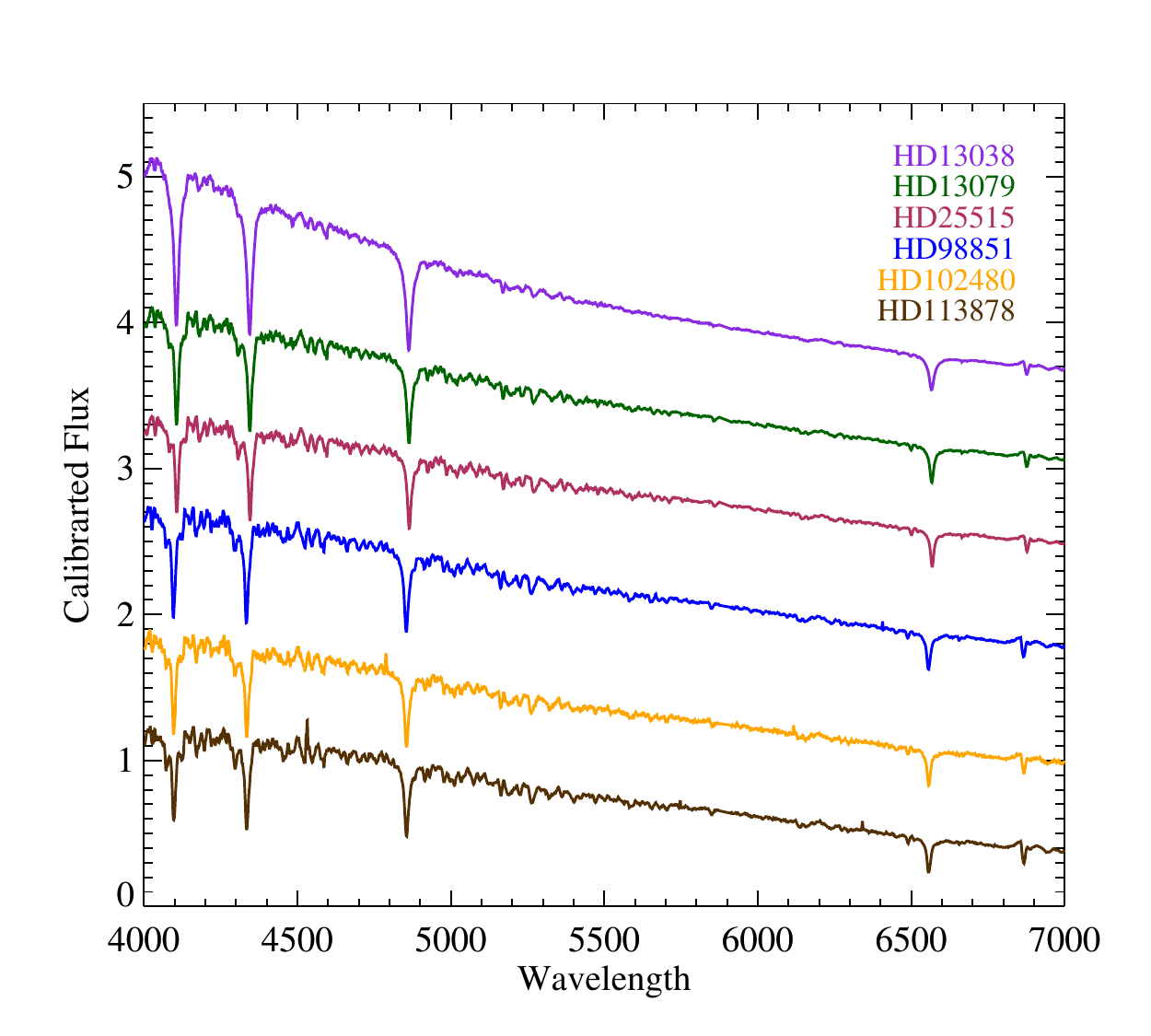}
  \caption{Low-resolution spectra of sample stars observed from HCT. Flux values have been rescaled to adjust the spectra for all the stars in the same frame.}
  \label{fig:sample_fig}
 \end{center}
\end{figure}

Low-resolution spectroscopic observations were carried out at Hanle Faint Object Spectrogrpah Camera (HFOSC) mounted on 2.0-m Himalayan Chandra Telescope (HCT), operated by Indian Institute of Astrophysics (IIA) at Hanle, Ladakh, India. We used SITe CCD of 2K$\times$4K pixels with pixel size of $15\mu\times15\mu$ and image scale of 0.296"/pixel. Spectroscopy on the HCT/HFOSC was done in the wavelength range of 4000-7000 \AA{} using GR7 and slit width of 0.77", giving a spectral resolution R $\sim$ 1300. The bias, flats, arcs were observed on each night for the calibration purpose. Spectroscopic data reduction was done under IRAF environment. The wavelength calibration was done using about 50 emission lines of Fe-Ar lamp with a typical uncertainty of 1.26 \AA. Feige 110 and Feige 34 were used as standards for the flux calibration. Final flux-calibrated spectra are shown in Fig. \ref{fig:sample_fig}.

\begin{figure}
 \begin{center}
  \includegraphics[scale=1.4]{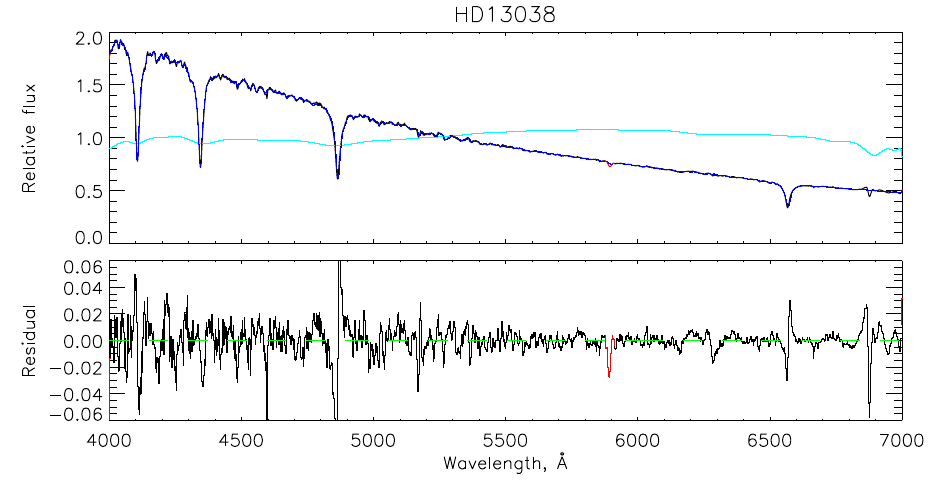}
  \caption{Full spectrum fitting for a sample star, HD 13038. Upper panel shows the observed spectrum in black and fitted one in blue. Cyan line shows the multiplicative polynomial. Lower panel shows the residuals between the two (Observed\,-\,fitted).}
  \label{fig:fitting_fig}
 \end{center}
\end{figure}

\section{Low resolution spectroscopic parameters}
\label{sec:parameters}

Atmospheric parameters have been derived using the method of full spectrum fitting technique based on $\chi^2$ minimization. For this purpose, we use IDL based ULySS package with MILES V2 `spectral interpolator' (Sharma et al., 2016). Role of spectral interpolator is to generate the model spectrum, based on a stellar spectral library (MILES, in our case) for the comparison.

Spectra were fitted in the wavelength range 4000\,-\,7000 \AA{} using the same approach as adopted in Sharma et al. (2016). Following values of the initial guesses were supplied in order to avoid the trapping of the solution in the local minimum valley: \teff{}~$\in [5000, 10000, 15000]$\,K, \logg{}~$\in [1.0, 4.0]$\,dex, and \feh{}~$\in [-2.0, -0.5, 0.3]$\,dex. Although, spectrophotometric standards were used for calibrating the fluxes of the sample stars while reducing the data, but to absorb any remnant flux, we used $40^{\scriptsize{\textrm{th}}}$ order of multiplicative polynomial.

\begin{table}
\caption{Derived atmospheric parameters for the sample stars. High resolution spectroscopic parameters are taken from Joshi et al. (2017).}
\label{tab:params}.
\centering
\scalebox{0.85}{
\begin{tabular}{clccrcc}
\hline
S. No. & Star   &  \multicolumn{3}{c}{Low\,-\,resolution spectroscopic parameters} & \multicolumn{2}{c}{High\,-\,resolution Parameters} \\
\hline   
   &            & \teff       & \logg         & \multicolumn{1}{c}{\feh}   & \teff       & \logg          \\
   &            &  (K)        &  (dex)        &  \multicolumn{1}{c}{(dex)} & (K)         &  (dex)        \\
\hline
1. & HD\,13038  & 8585$\pm$365   &  4.36$\pm$0.29    &  0.07$\pm$0.30      & 7960$\pm$200 & 3.80$\pm$0.15 \\
2. & HD\,13079  & 7426$\pm$185   &  4.07$\pm$0.45    &  0.27$\pm$0.30      & 7040$\pm$200 & 3.40$\pm$0.20 \\
3. & HD\,25515  & 7110$\pm$162   &  3.99$\pm$0.50    &  0.31$\pm$0.30      & 6650$\pm$200 & 3.80$\pm$0.15 \\
4. & HD\,98851  & 7269$\pm$122   &  3.97$\pm$0.33    &  0.44$\pm$0.30      & 7000$\pm$200 & 3.65$\pm$0.15 \\
5. & HD\,102480 & 7106$\pm$149   &  3.95$\pm$0.47    &  0.27$\pm$0.30      & 6720$\pm$200 & 2.90$\pm$0.20 \\
6. & HD\,113878 & 7138$\pm$173   &  3.81$\pm$0.57    &  0.55$\pm$0.30      & 7000$\pm$200 & 3.70$\pm$0.10 \\
\hline                                                                                                   
\\
\end{tabular}
}
\end{table}

Derived atmospheric parameters are listed in Table~\ref{tab:params}. Errors reported are the internal errors (returned by ULySS) rescaled by a scaling factor. For \teff{} and \logg{}, these factors are of the order of 10 while for \feh, the scale factor is of the order of 30.

\section{Discussion and future plans}\label{sec:discussion}

These stars have recently been studied by Joshi et al. (2017) in high-resolution spectroscopic mode. Comparison of two series of measurements shows an average difference of $-377\pm178$\,K, and $-0.48\pm0.32$\,dex for \teff{} and \logg{} respectively, whereas the mean estimated errors are 193\,K and 0.43\,dex for the two parameters. Within the uncertainties, the two series of measurements are consistent.

Full spectrum fitting method has been employed on peculiar stars for the first time, therefore it is important to validate the method for a larger set of such objects. For this purpose, we plan to enhance our test sample by gathering the CP stars' spectra from various spectral archives. Once validated, this method would be useful as a parameter reduction pipeline for the low resolution spectra obtained using FOSC and other future instruments at 3.6-m Devasthal Optical Telescope (DOT).
%
%
\section*{Acknowledgements}

KS thanks Neelam Panwar for assistantship through DST-INSPIRE faculty award. 
This research has made use of the SIMBAD database and VizieR catalogue tools, operated at CDS, Strasbourg, France.
%
%
%

\footnotesize
\beginrefer

\refer Catanzaro G., Ripepi V., 2014, MNRAS, 441, 1669

\refer Joshi S., et al., 2003, MNRAS, 344, 431

\refer Joshi S., et al., 2012, MNRAS, 424, 2002

\refer Joshi S., Semenko E., Moiseeva A., Sharma K., et al., 2017, MNRAS

\refer Koleva M., Prugniel P., Bouchard A., Wu Y., 2009, A\&A, 501, 1269

\refer Lipski {\L}., St{\c e}pie{\'n} K., 2008, MNRAS, 385, 481

\refer Martinez P., et al., 2001, A\&A, 371, 1048

\refer Perryman, M. A. C., de Boer, K. S., Gilmore, G., et al. 2001, A\&A, 369, 339

\refer Prugniel, P., Vauglin, I., \& Koleva, M. 2011, A\&A, 531, A165

\refer Sharma K., Prugniel P., Singh H. P., 2016, A\&A, 585, A64

\refer Wu Y., Singh H. P., Prugniel P., Gupta R., Koleva M., 2011, A\&A, 525, A71

\refer York, D. G., Adelman, J., Anderson, Jr., J. E., et al. 2000, The Astronomical Journal, 120, 1579

\endrefer           


\end{document}